\documentclass[aps,prl,superscriptaddress,unsortedaddress,twocolumn]{revtex4}
\usepackage{amsmath,epsfig,mathrsfs,graphicx,amssymb,slashed}

\begin{document}
\title{Local Realism in Quantum Many Worlds}
\author{Adam Bednorz}
\email{Adam.Bednorz@fuw.edu.pl}
\affiliation{Faculty of Physics, University of Warsaw, ul. Pasteura 5, PL02-093 Warsaw, Poland}
%\date{\today}

\begin{abstract}
  Fundamental principle of classical physics -- local realism, means that freely chosen  observations can be explained by  a local (slower than light) real process. It is apparently violated in quantum mechanics as shown by Bell theorem. Despite extreme efforts  experiments have not conclusively confirmed this violation due to loopholes.
We propose a new postulate that the description of quantum processes must be consistent with local realism,
It also assumes existence of many worlds/copies of the same system, interacting  weakly microscopically but strongly macroscopically, whose number can be estimated experimentally.Bell theorem will never address a real experiment because its assumptions cannot be strictly fulfilled. By an appropriate generalization of quantum framework and measurement postulates, in particular taking into account freedom of choice, local realism agrees with quantum mechanics
and the performed experiments, also involving single qubit coherence  and a weaker version of the Bell test, Einstein-Podolsky-Rosen  steering. 
\end{abstract}

\maketitle 

Local realism means that we can reproduce results of an experiment by some real process with information transfer slower than light, subluminal. The question of local realism makes sense only if we agree on \emph{freedom of choice}, otherwise all experiments can be explained by superdeterminism (correlations from the far past). Then the choice creates some information to be transferred locally (subluminally). Local realism is trivally correct in classical mechanics, because it itself provides the desired local process. This is no longer obvious in quantum mechanics because the all we have are detection results with no direct construction of the process. Even worse, Bell theorem 
\cite{bell} states that it is impossible for a special entangled state where appropriate choice of measurements by two remote observers violates some inequality satisfied by all local realistic explanations \cite{chsh}.
Many experiments confirmed this violation 
\cite{belx1a,belx1b,belx1c,belx1d,belx1e,belx1f,belx1g,belx1h,belx2a,belx2b} but always with at least one loophole.
The loophole means that some assumption of the Bell test has been weakened, either the detector is not sufficiently efficient (most of detected particles are lost), the distance between observers is too small to exclude local communication or there is no free choice. Extreme efforts are taken to close all loopholes simultaneously,
which some experts speculate to be achieved soon \cite{kwgi}.

The absence of a loophole-free violation of Bell-type inequalities admits a local and real process exploring loopholes \cite{loop1,loop2,loopmod1},  supported
by more or less realistic models \cite{loopmod1,loopmod2,loopmod3,loopth}. 
 A large class of realistic models have been excluded by realization of Einstein-Podolsky-Rosen (EPR) paradox \cite{epr}.
It is tested by \emph{steering} --  a weaker version of the Bell test where one of the observers uses agreed quantum description. It has been confirmed loophole-free experimentally \cite{eprst1,eprst2,eprst3}.  Nevertheless, a general construction of
a model of local reality consistent with quantum mechanics has never been precluded. The Bell theorem \cite{bell,chsh} itself relies on
assumptions about quantum measurement: (A) an ideal entangled state without decoherence, (B) instant and perfect choice and measurement. Although these assumptions are allowed by standard theory of quantum measurement, they do not need to be feasible in reality. Here we show that such construction is possible within quantum framework with all choices potentially simultaneously measurable. This construction almost surely will involve many worlds (copies of a single one) which however do not split constantly \cite{mwi} (it would be nonlocal) but with a number possibly large but fixed and interacting locally and weakly \cite{mww}.

A general construction and restriction of quantum observations, satisfying principle of local realism, will be completed if the observations  depend locally on free-to-choose options, readouts for all options simultaneously are represented by a positive probability.
All events, free choices and measurements will be referred by time-position $x=(x^0=ct,\vec{x})$
(time $t$, spatial coordinates $\vec{x}$, speed of light $c$).
Given the initial state of the system (universe) and it dynamics, Hamiltonian $H$,
the \emph{free choice} $a$ means that a decision whether to modify the dynamics by an extra term in the Hamiltonian $H_a(x)$ localized at $x_a$. This corresponds to the usual Bell-type experimental situation of a dichotomic choice. There can be many such defined choices localized at the points, $a,b,c,...$. 
The readout cannot depend on remote choice, beyond causality region.
The original Hamiltonian with or without extra free terms satisfies causality, any change cannot propagate faster than light. Every choice $a$ results in different dynamics within the future lightcone starting at $a$, $x^0-x^0_a\geq |\vec{x}-\vec{x}_a|$. In overlapping lightcones of $a$ and $b$ the dynamics depends on joint choices.  Let $O(x)$ denote an observable in Heisenberg picture with respect to the original Hamiltonian, while for $O_a(x)$, $O_b(x)$, $O_{ab}(x)$ we add the choice-dependent Hamiltonian $H_a$, $H_b$, or $H_a+H_b$, respectively. Certainly the causality principle tell us that
if $x$ is outside of the future lightcone of $x_a$ then $O_a(x)=O(x)$ and $O_{ab}(x)=O_b(x)$. If $x$ is outside both future lightcones of $x_a$ and $x_b$ then $O_{ab}(x)=O(x)$. 

\begin{figure}
\includegraphics[scale=.4]{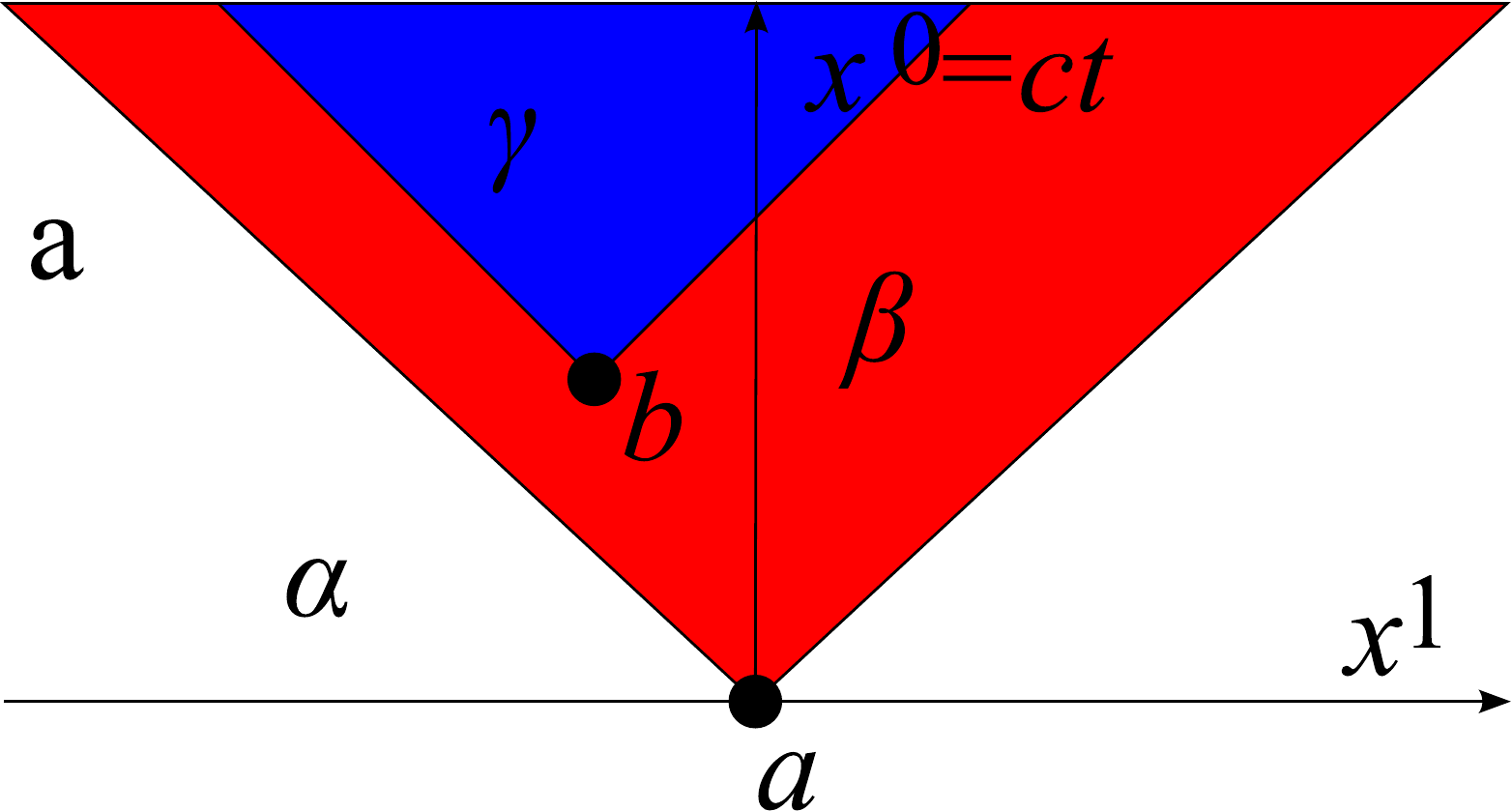}
\includegraphics[scale=.4]{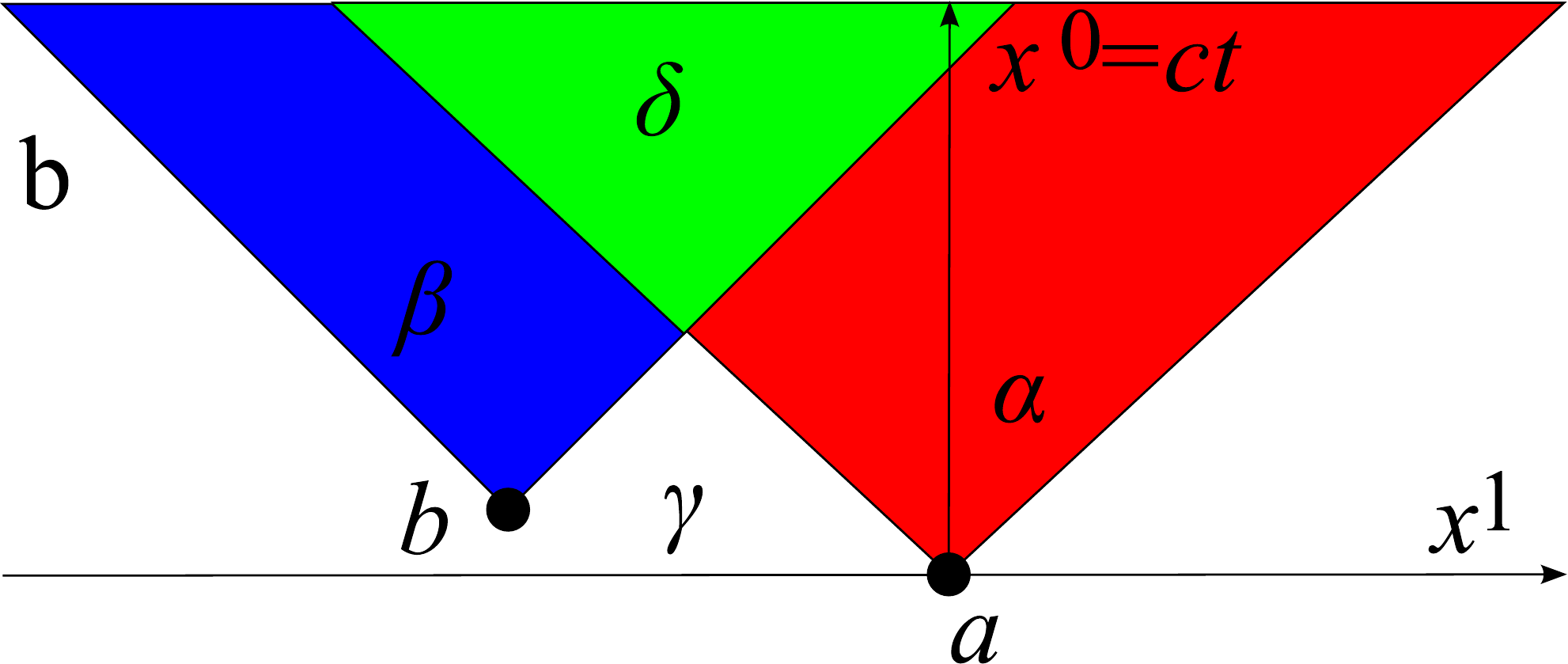}
\caption{Examples of free choices and readout spatiotemporal configuration (time $x^0=ct$ and position $x^1$).
(a) Choice $b$ lies inside the lightcone of $a$ dividing the spacetime into $3$ readout regions: $\alpha$ independent of $a$ and $b$,
$\beta$ depending on $a$ but not $b$, $\gamma$ depending on $a$ and $b$.
(b) Choices $a$ and $b$ are outside of each other's lightcone so there are $4$ regions: $\gamma$ independent of $a$ and $b$, $\alpha$ depending on $a$ but not $b$, $\beta$ depending on $b$ but not $a$, $\delta$ depending on $a$ and $b$.}
\label{spatio}
\end{figure}

The  measurement readout which will be also localized in spacetime, i.e. a random function $\alpha(x)$ (or $\beta(x)$ etc.). The probability is given by positive operator-valued measure (POVM) \cite{povm}, as
$\mathrm{Tr}K\rho K^\dag$ with the Kraus operator $K$ in the state $\rho$ ($=|\psi\rangle\langle\psi|$ for a pure state). To satisfy causality $K$ can depend only on the causal past of $x$, in Heisenberg picture. 
The main point of this work is to impose the condition of local realism already upon Kraus operators -- a single POVM for all choices simultaneously. Namely, all readouts will be choice-conditioned (only for these choices that can affect readout) , e.g. $\alpha\to \alpha,\alpha_a,\alpha_b,\alpha_{ab}$. This means that readouts for all choices, also those not just realized, are measurable. The readout points ($x_\alpha$, $x_\beta$, etc.) can lie within the future lightcone of some (but not necessarily all) points, as depicted in Fig. \ref{spatio}. Then the Kraus operators for a local realistic POVM must have the form
\begin{equation}
K[\{\alpha\},\{\beta\},\{\gamma\},...]
\end{equation}
where the set $\{\alpha\}$ denotes values for each combination of  \emph{only those choices that can influence} $\alpha$, i.e. $x_\alpha$ lies within their future lightcones. In the simple case of Fig.\ref{spatio}a, the choice $b$
 lies inside the future lightcone of $a$, so that the full Kraus operator reads
\begin{equation}
K(\alpha,\beta,\beta_a,\gamma,\gamma_a,\gamma_b,\gamma_{ab})
\end{equation}
In the situation in Fig.\ref{spatio}b, $x_a$, $x_b$ lie outside of each other future lightcone and then the maximal dependence of Kraus operator reads
\begin{equation}
K(\gamma,\alpha,\alpha_{a},\beta,\beta_{b},\delta,\delta_a,\delta_b,\delta_{ab})
\end{equation}

\begin{figure}
\includegraphics[scale=.3]{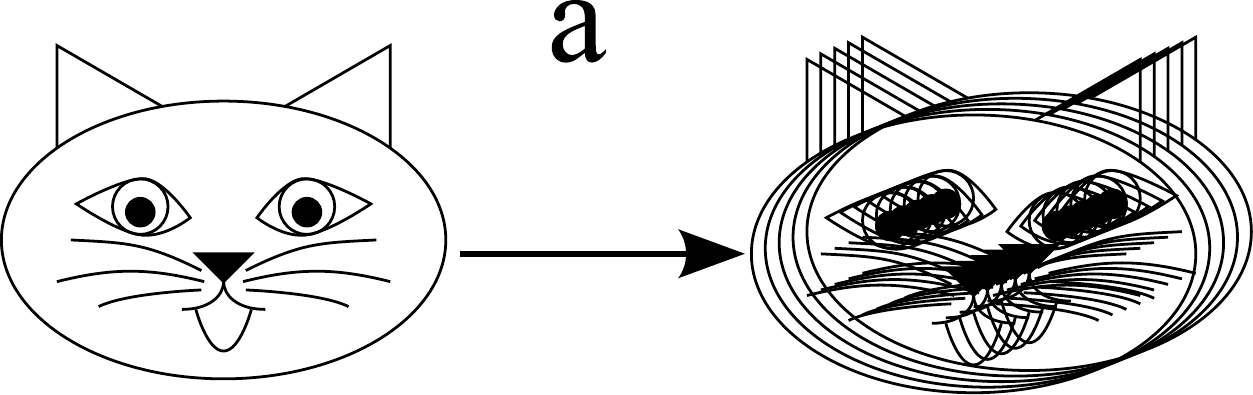}
\includegraphics[scale=.3]{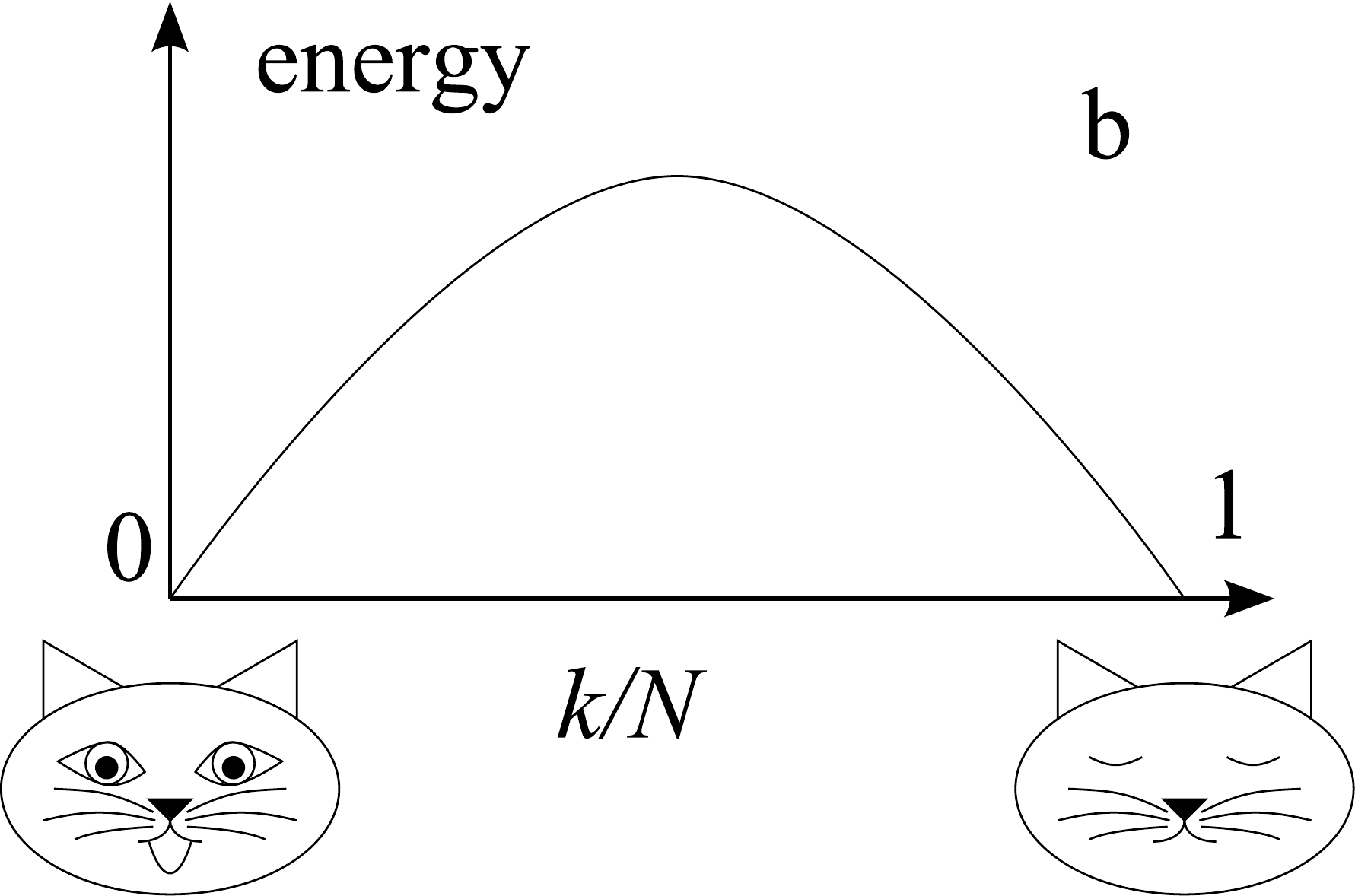}
\caption{Schr{\"o}dinger cat in many worlds. (a) A cat we see is in fact a bunch of copycats. (b) The  macrostate with only part of copycats dead/alive is energetically unstable.}
\label{catt}
\end{figure}

Let us demonstrate how the joint POVM works in the simplest and most studied cases.
An evolving qubit -- single two level system (e.g. spin, ion, atom) can be prepared in some state, e.g. $|+\rangle$  and evolve by a Hamiltonian $H=\hbar\omega|+\rangle\langle -|+$h.c. so that the state rotates in time  $t$ into
$\cos\omega t|+\rangle-i\sin\omega t|-\rangle$. Suppose we want to measure if we have the initial state after time $t$. This corresponds to the recent experimental situation  with long coherent ion qubits \cite{ions} but also artificial qubits \cite{arti1,arti2,arti3} (though of much shorter coherence). Using projection $P=|+\rangle\langle +|$ one finds the probability
$p_t=\cos^2(\omega t/2)$. Now suppose that we can choose to read it or not at two times $t_a$ and $t_b$ which corresponds
to the situation in Fig. \ref{spatio}a. In particular the interesting values will be $\beta_a$ and $\gamma_b$ which will be $1$ if the state was initial and $0$ otherwise. We have to find  at least joint probability $p(\beta_a,\gamma_b)$. Of course such a probability, reproducing projective predictions, exists e.g. $p=p_ap_b$, but we have to construct it using POVM. A simple attempt,
$K(\beta_a=\gamma_b=1)=P_bP_a$, results in disturbance of the last readout by the first. If $t_b=\pi/\omega$ while $t_a=\pi/2\omega$
then $p=1/4$ for every event, in conflict with projective expectation $p(\gamma_b=1)=p(1,0)+p(0,0)=0$.
A two-state representation of the qubit fails. Suppose that there exist many copies or worlds of the same qubit.  Then the actual state of the qubit $c|+\rangle+d|-\rangle$ can be
$\prod_{j=1}^N(c|+\rangle+d|-\rangle)_j$ with the Hamiltonian $H=\hbar\omega\sum_j|+\rangle_j\langle -|+$h.c.
The POVM can just pick two (or more) different copies $K=P_{bj}P_{ak}$ with $P_j=|+\rangle_j\langle +|$ and $j\neq k$.
Importantly, 
the worlds are not splitting at each measurement act \cite{mwi} -- this would be manifestly nonlocal. Their number is constant, they occupy \emph{the same} spacetime and there may be some weak \emph{local} interaction \cite{mww}
that makes the worlds similar at macroscopic level, but not for single isolated qubits. It essentially means that the original Hilbert space and Hamiltonian is copied $N$ times
$H\to\sum_j H_j$ with additional very weak inter-world interaction \cite{mww,plag} relevant only at macroscale.
For two choices $N=2$ suffices but in general $N$ must match at least the number of choices. Such an estimate should be possible to find in ion or artificial qubits experiments \cite{ions,arti1,arti2,arti3}.
The fact that we see a single world macrosopically, just like the Schr{\"o}dinger cat is alive or dead but not both can  result from weak inter-world interaction. Promoting the qubit to become a dead/alive macroscopic state of a cat, the energy can depend on the fraction of $|-\rangle$ in the state, namely $|+\rangle^{N-k}|-\rangle^{k}$ has energy $E_k$ with the minimum $0$ only for $k$ equal $0$ or $N$ --
when the state in all worlds is the same. Since other macrostates are energetically unfavorable, the system will ultimately \emph{collapse} to either $|+\rangle^N$ or $|-\rangle^N$, see. Fig. \ref{catt}.

The problem of insufficient Hilbert space occurs also in the local realistic description of EPR-steering.
The EPR-steering experiments \cite{eprst1,eprst2,eprst3} violate local realism under a stronger assumption than Bell.  Of the two separate observers, one of them (Bob) \emph{trusts} in the particular quantum representation of his party of the 
measured system and detection. Bob can freely choose 
between two, there or more measurements with dichotomic readout, $\pm 1$. Bob in fact can also read $0$
out, but these events are discarded. Alice, on the contrary, does not assume anything about 
her POVM and her possible readouts are $\pm 1$ and $0$, without discarding anything. Bob \emph{assumes} that his part can be represented by a qubit, with the basis states $|+\rangle$ and $|-\rangle$, and the choice corresponds to some unit vector on Bloch sphere, $\vec{b}$, $|\vec{b}|=1$. The corresponding, trusted POVM for the readout $\pm 1$ at Bob's side reads 
$K_B(\pm 1)=(1\pm\vec{b}\cdot\vec{\sigma})_B/2$, with the Pauli matrices 
$\vec{\sigma}=(|-\rangle\langle +|+|+\rangle\langle -|,i|-\rangle\langle +|-i|+\rangle\langle -|,|+\rangle\langle +|-|-\rangle\langle -|)$. Suppose that Bob (and Alice) can choose between $\vec{b}_1$, $\vec{b}_2$, $\vec{b}_3$, or more.
Then our construction implies a joint Kraus operator for all choices $K(b_1,b_2,b_3,a_1,a_2,a_3)$ for $b_j=\pm 1$, $a_j=\pm 1,0$ which reduces to the
trusted Bob's POVM when ignoring not chosen options, i.e.
\begin{equation}
\sum_{a_2,a_3,b_2,b_3} K^\dag K=(K^\dag K)_A(a_1)(K^\dag K)_B(b_1)
\end{equation}
where $K_{A/B}$ applies to the Alice/Bob subspace and $K_B$ is given above.  Alice can also make two or three choices.
The EPR-steering paradox appears if one compares possible results from a local realistic POVM, joint for all choices,
as described above, with \emph{separate} POVMs for different choices. Suppose Alice makes has the same choice as Bob and her readouts correspond to the same set of Kraus operators. If Alice and Bob share a maximally entangled state
$(|+-\rangle-|-+\rangle)/\sqrt{2}$ then the ideal joint probability reads $p(a,b)=(1-\vec{a}\cdot\vec{b})/4$ ($0$ is not registered 
by Alice) where $a=\pm 1$ corresponds to the vector $\pm\vec{a}$. if $\vec{a}=-\vec{b}$ then the correlation is perfect, 
namely $p(a=b)=1/2$ and $\langle ab\rangle=1$. This theoretical result is incompatible with local realistic POVM (with 
trusted Bob's readouts) as it violates certain inequalities, by inspection of perfect correlations. A simple attempt would be Bob's POVM as above for a random directions of the set of choices. Only if the actual choice match then the readout is registered, otherwise discarded. Alice does the same, but her $0$ must be included in the correlations leading to $1/M$ suppression of $\langle ab\rangle$.

For $M=3$ and orthogonal Bob/Alice directions: $\vec{b}_1=(1,0,0)$, $\vec{b}_2=(0,1,0)$, $\vec{b}_3=(0,0,1)$ the following inequality holds for any local realistic (Bob-trusted) POVM \cite{eprst1},
\begin{equation}
T=\sum_{j=1,2,3}\sum_{a_j}p(a_j)\langle b_j\rangle^2_{a_j}\leq 1/3\label{eprin}
\end{equation}
where $p(a_j)$ is the probability of occurring $a_j$  when the choice $j$ is made by Alice while $\langle b_j\rangle_{a_j}$ is the conditional average of $b_j$ for a particular value $a_j$ occurred and Bob also chose $j$.
Even with 
experimental defficiences, the violation has been confirmed  \cite{eprst1,eprst2,eprst3}), satisfying other Bell conditions (free choice and no communication).

To resolve the conflict one has to relax the Bob's trust. 
Again, we multiply the initial space. Alice and Bob do not share a \emph{single} entangled 
state but their \emph{many copies} so that the full state is a tensor product $\prod_k(|+-\rangle-|-+\rangle)/\sqrt{2}$
for  $k=1,...,N$. Let Bob choose the direction $\vec{b}$ and measure every copy of his qubit with $K_B(b^k)$ for the value $b^k=\pm 1$. Now the reported readout is $+1$ only if $b^k=+1$ for \emph{every} $k$ and $-1$ if $b^k=-1$ for every $k$. If not all $b^k$ are equal then the reported readout is $0$ and the event, regardless the Alice's readout, is discarded. For several choices, the actual direction in $K$ is taken randomly from the set of all possibilities. If it matches the actual choice, then the readout according the above scheme is reported, otherwise discarded. Then Alice can construct local realistic Kraus operator similarly as in the case of free qubit, $K_A=\prod_k K_A(a_k)$ where $K_A(a_k)$ acts only on the $k$th qubit.
Then the correlations are perfect and EPR-steering is consistent with local realism. EPR-steering experiments can estimate the lower bound for the number of copies, which must be at least 2-3, according to the present data \cite{eprst1}.

\begin{figure}
\includegraphics[scale=1.5]{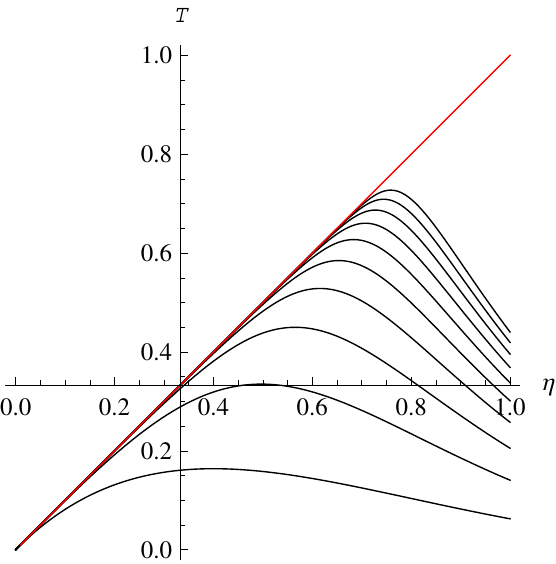}
\includegraphics[scale=1.5]{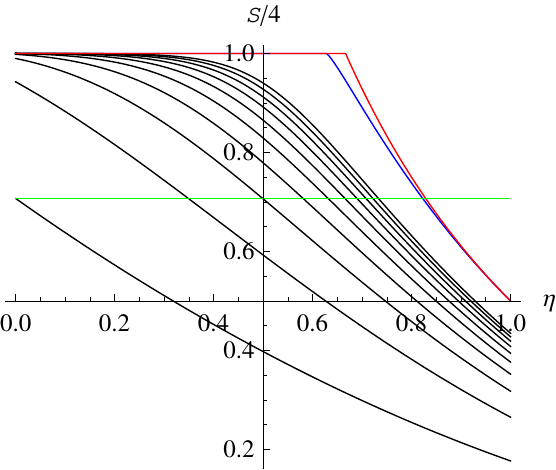}

\caption{Steering and Bell parameter, $T$ and $S$, respectively with respect to efficiency for local realistic POVMs for $N=1$ to $10$ copies of the entangled state (black, $N$ increases with the height of the curve) and $N\to\infty$ (blue\cite{loopth}). Maximal bound for $T$ and $S$ in local realism corresponds the  red line.}\label{chob}
\end{figure}

The final case is the Bell test, which gives the same correlations as in EPR-steering, $\langle ab\rangle=-\vec{a}\cdot\vec{b}$ and $a,b=\pm 1$, but a general POVM is allowed (no trust from either Alice's or Bob's side). Taking two choices for Alice and Bob, $\vec{a}_1=(1,0,0)$, $\vec{a}_2=(0,1,0)$, $\sqrt{2}\vec{b}_1=(-1,-1,0)$, $\sqrt{2}\vec{b}_2=(-1,1,0)$ then
$S=\langle a_1b_1\rangle+\langle a_1b_2\rangle+\langle a_2b_1\rangle-\langle a_2b_2\rangle$ gives $2\sqrt{2}$
in conflict with local realistic bound $2$.\cite{chsh} However, in contrast to EPR-steering, this violation has not been confirmed experimentally \cite{belx1a,belx1b,belx1c,belx1d,belx1e,belx1f,belx1g,belx1h,belx2a,belx2b} without loopholes \cite{loop1,loop2,loopmod1}. Restricting to those experiments that satisfied spacelike separation during the measurement the major loophole is low efficiency, $\eta=p(a^2=b^2=1)/p(a^2=1)$, taking into account that $a,b$ can be sometimes $0$. We can easily construct local realistic POVM with $\eta=50\%$. To this end, we pick randomly one of the directions $\vec{a}_j$, $j=1,2$ for Alice. If it matches the actual choice then the value according to $K_A(a_j)$ is reported and otherwise $0$ (hence $50\%$), and similarly for Bob. In this case the \emph{coincidence} correlation $\langle ab\rangle_{a^2=b^2=1}=-\vec{a}\cdot\vec{b}$ and only the efficiency is bounded by $50\%$. Although at present no experiment reported such high efficiency \emph{in conjunction} with other Bell conditions, one can push the limit higher (the absolute bound is $2(\sqrt{2}-1)\simeq 83\%$ \cite{loopmod1}) with help of $N$ copies/worlds of the entangled state. Let Alice and Bob preselect states with total spin $\vec{J}_{A/B}=\sum_j\vec{\sigma}_{A/B}/2$ and make spin tomography, namely take $(N+1)^{1/2}K_{\vec{A}}=P_{\vec{A}}$ with the uniform measure on unit sphere $|\vec{A}|=1$. Here
$P_{\vec{A}}=|\vec{A}\rangle\langle\vec{A}|$ is projection for
$J_{\vec{A}}=\vec{A}\cdot\vec{J}$ and $J_{\vec{A}}|\vec{A}\rangle=(N/2)|\vec{A}\rangle$, so that $|\vec{A}\rangle$ is 
a spin coherent state \cite{cohe} and $p=\mathrm{Tr}K^\dag K\rho$. This preselects states
with the eigenvalue (joint for $A$ and $B$) $\vec{J}^2=N(N/2+1)/2$ with the overall probability $(N+1)/2^{N}$ and the rest are discarded (zero).
For this ensemble the coherent state tomography results in the probability distribution
\begin{equation}
p(\vec{A},\vec{B})=\frac{1}{(4\pi)^2(N+1)}\left(\frac{1-\vec{A}\cdot\vec{B}}{2}\right)^N
\end{equation}
which becomes $\delta(\vec{A}-\vec{B})/4\pi$  \cite{loopth} in the limit $N\to\infty$.
Now, Alice and Bob take their directions $\vec{a},\vec{b}$, some fixed number $q\in [0,1]$ and calculate 
$\vec{a}\cdot\vec{A}$ or $\vec{b}\cdot\vec{B}$. They assign the final readout: $+1$ if $> +q$, $0$ for $[-q,+q]$, $-1$ 
if $<-q$. For the same angles as in usual EPR steering and Bell test, it leads to apparent violation of local realism with efficiency close to 
the absolute bound in the limit $N\to\infty$ when it reduces to chaotic ball \cite{loopth}. The violation is 
already considerable for relatively small $N$ as depicted in Fig. \ref{chob}.

In conclusion, quantum mechanics may have some strange nonclassical features such as violation of time-reversal symmetry \cite{asym} but not necessarily violates local realism.
We have demonstrated that the latter can be reconciled with quantum mechanics by making two amendments in the theoretical description: (A) Joint measurement description for all choices simultaneously, (B) introduction of many worlds -- the actual system is multirepresented. Contrary to the original idea, the worlds are not splitting (their number is constant) and interact locally and weakly microscopially but strongly marcoscopically making them similar in the observable reality. Further experiments are needed to confirm this conjecture, especially to estimate possible inter-world interaction. One has to be also cautious about soon possible announcement of a loophole-free Bell test, as time of choice and readout should be in principle not machine but human-concluded \cite{wise}, rather a long-term perspective.

W. Belzig and M.W. Hall are acknowledged for motivation and discussion.

\end{document}